\documentclass[reprint,
superscriptaddress,
 amsmath,amssymb,
 aps,
]{revtex4-2}

\usepackage{graphicx}
\usepackage{dcolumn}
\usepackage{bm}
\usepackage{color}

\usepackage{color}

\begin{document}

\preprint{APS/123-QED}

\title{Anisotropic superconducting gap probed by $^{125}$Te NMR in noncentrosymmetric Sc$_6M$Te$_2$ ($M$ = Fe, Co)}

\author{Kanako Doi}
\affiliation{Department of Physics, Nagoya University, Chikusa, Nagoya 464-8602, Japan.}
\author{Hayase Takei}
\affiliation{Department of Physics, Nagoya University, Chikusa, Nagoya 464-8602, Japan.}
\author{Yusaku Shinoda}
\affiliation{%
 Department of Applied Physics, Nagoya University, Chikusa, Nagoya 464-8603, Japan.}%
\author{Yoshihiko Okamoto}
\affiliation{%
 Institute for Solid State Physics, University of Tokyo, Kashiwa, 277-8581, Japan.
}%
\author{Daigorou Hirai}
\affiliation{%
 Department of Applied Physics, Nagoya University, Chikusa, Nagoya 464-8603, Japan.}%
\author{Koshi Takenaka}
\affiliation{%
 Department of Applied Physics, Nagoya University, Chikusa, Nagoya 464-8603, Japan.}%
 \author{Taku Matsushita}%
\affiliation{Department of Physics, Nagoya University, Chikusa, Nagoya 464-8602, Japan.}
\author{Yoshiaki Kobayashi}%
\affiliation{Department of Physics, Nagoya University, Chikusa, Nagoya 464-8602, Japan.}
\author{Yasuhiro Shimizu}%
\affiliation{Department of Physics, Nagoya University, Chikusa, Nagoya 464-8602, Japan.}
\affiliation{Department of Physics, Shizuoka University, Suruga, Shizuoka 422-8529, Japan.}
 
\date{\today}

\begin{abstract}
The superconducting gap symmetry is investigated by $^{125}$Te NMR measurements on Sc$_6M$Te$_2$ ($M$ = Fe, Co) without spatial inversion symmetry. The spin susceptibility obtained from the Knight shift $K$ is suppressed below the superconducting transition temperature, while leaving a finite value down to the lowest temperature ($\simeq 0.4$ K). The nuclear spin-lattice relaxation rate $1/T_1$ follows a power law against temperature $T$ without showing a coherence peak characteristic of the isotropic gap. The result implies a pairing admixture or a residual density of states under magnetic field. The normal metallic state has a Korringa scaling relation between $1/T_1T$ and the Knight shift, reflecting a weak electron correlation.
\end{abstract}

\maketitle


\section{Introduction}
Noncentrosymmetric superconductors have attracted attention due to parity mixing and reciprocal transport \cite{Bauer2012, Smidman2017, S.Yip2014noncentrosymmetricSC, Wakatsuki2017, Tokura2018}. The antisymmetric spin-orbit coupling under strong spin-orbit coupling and broken spatial inversion brings out unconventional Cooper pairing with a mixed parity of even spin-singlet and odd spin-triplet symmetry \cite{Frigeri2004PRL}. Topological superconductors including a spin-triplet superconductor with Majorana bound states provide a significant platform for topological quantum computation that implements error corrections against local perturbation \cite{Sato2016}. The hallmark experimental evidence for spin-triplet pairing satisfies anisotropic spin susceptibility, enhanced Pauli-limiting field, zero-bias conductance, and time-reversal symmetry breaking \cite{Gorkov2001, Fujimoto2007, Mandal2023}. Noncentrosymmetric superconductors have been studied extensively in rare-earth metals such as CePt$_3$Si \cite{Bauer2004CePt3Si, Bonalde2005, Frigeri2004PRL, Samokhin2004}, transition-metal alloys \cite{Togano2004,Lee2005,Hillier2009,Bao2015,Barker2015,Shang2020}, topological semimetals \cite{Ali}, and artificial superlattice \cite{Ando2020}. The parity admixture strongly depends on electron correlation and spin-orbit coupling \cite{Fujimoto2007}. Thus, a physical and chemical tuning of these parameters is desired to quantitatively uncover the effect of antisymmetric spin-orbit coupling on superconducting pairing symmetry. 

Ternary scandium tellurides Sc$_6M$Te$_2$ ($M$: transition metals) have a hexagonal
Zr$_6$CoAl$_2$-type structure without inversion symmetry \cite{Chen2004Sc6TTe2, Maggard2000}. Superconductivity was discovered in Sc$_6$FeTe$_2$ with a hexagonal $P\bar{6}2m$ ($D^3_{3h}$, No.189) lattice at the highest transition temperature $T_c$ = 4.7 K among a series of materials \cite{Shinoda2023}. The upper critical field extrapolated to 0 K reaches 8.7 T, which is comparable to the Pauli paramagnetic limit. The second highest $T_c$ = 3.6 K has been observed in Sc$_6$CoTe$_2$ having the same lattice symmetry \cite{Shinoda2023}. Although the band splitting due to spin-orbit coupling and inversion symmetry breaking is smaller than that of rare-earth compounds, the strength of the electron correlation can be systematically tuned by a chemical substitution of the $M$ site with $3d$, $4d$, and $5d$ transition metals. The density of states $D(E_{\rm F})$ at the Fermi level $E_{\rm F}$ is mainly composed of Sc and $M$ bands in the band calculation. $E_{\rm F}$ is located close to a peak of $D(E)$ for $M$ = Fe and a valley for Co. The Sommerfeld coefficient obtained from the specific heat, $\gamma$ = 73 (Fe) and 55 (Co) mJ/K$^2$mol, is significantly greater than the calculated free electron value $\gamma \sim 22$ mJ/K$^2$mol, suggesting a strong electron correlation. The specific heat jump ($\Delta C/\gamma T_c$ = 2.4) in Sc$_6$FeTe$_2$ at $T_c$ well exceeds the value ($\Delta C/\gamma T_c$ = 1.43) expected for the weak-coupling Bardeen-Cooper-Schrieffer (BCS) superconductor and thus implies strong-coupling superconductivity. 

We utilize NMR spectroscopy as a powerful probe of parity mixing in noncentrosymmetric superconductors with strong spin-orbit coupling \cite{Hayashi2006, Nishiyama2005, Nishiyama2007, Yanase2008, Mukuda2008, Kandatsu2008, Matano2013, Matano2014SrPtAs, Nagase2020,Yogi2004}. The Knight shift, which measures the local spin susceptibility at the selected atomic site, distinguishes the predominant contribution from the spin-singlet and spin-triplet pairing. Since spin is not the conservative quantity under spin-orbit coupling, the possibility of the spin-triplet admixture state must be carefully examined from the pseudo-spin susceptibility \cite{Frigeri2004NJ}. The nuclear spin-lattice relaxation rate $1/T_1$ as a function of temperature measures the coherence factor of superconducting pairing and thus measures the spatial symmetry of the order parameter.

In this paper, we investigate the local spin susceptibility of superconducting and normal states through $^{125}$Te NMR measurements on Sc$_6$FeTe$_2$ and Sc$_6$CoTe$_2$ without inversion symmetry. The NMR spectrum uncovers the absence or presence of the structural symmetry-breaking transition. The results of Knight shift and $1/T_1T$ are compared with the first-principles band calculation. We examine the possible pairing symmetry and the effect of electron correlation. 

\section{Experimental}
$^{125}$Te NMR measurements were conducted on polycrystalline samples of Sc$_6M$Te$_2$ ($M$ = Fe, Co) synthesized with the arc melting method under Ar atmosphere \cite{Maggard2000, Chen2004Sc6TTe2, Shinoda2023}. The sample was dispersed in melted paraffin to avoid eddy current under NMR radio frequency (rf) pulses. Instead of the $^{45}$Sc nuclear spin ($^{45}I$ = 7/2) with a complex quadrupole-split NMR spectrum, here we performed $^{125}$Te ($^{125}I$ = 1/2) NMR with a simple spectrum. $T_{\rm c}$ under magnetic field was determined from a diamagnetic response of the $in$ $situ$ tank circuit and the intensity of the NMR signal. The NMR spectrum was measured under static magnetic fields of 2.57 and 9.13 T in the superconducting and normal states, respectively. The NMR spectrum was obtained from the Fourier transformation of the spin-echo signal taken after the rf pulses with the interval time $\tau = 30$ $\mu$s and the duration $t_{\pi/2}$ = 2 $\mu$s. $^{125}$Te Knight shift defined by $K = (\omega -\omega_0 )/ \omega_0 $ from the resonance frequency $\omega$ using the reference frequency $\omega_0 = \gamma_n \textit{H}$, where the bare gyromagnetic ratio $\gamma_n /2\pi =13.454$ MHz/T. The rf power dependence of the Knight shift was measured at 1.4 K, keeping the pulse length constant. We find no power dependence between 100 and 6 mJ, which provides evidence for the negligible heating effect below $T_c$. $1/T_1$ was measured with a saturation recovery method in which the nuclear magnetization recovery follows a single-exponential function above $T_c$ and becomes a stretched exponential with exponent $\beta$ = 0.8--1 below $T_c$. 
 
\section{Experimental results and discussion}
\begin{figure}
\includegraphics[width=8cm,keepaspectratio]{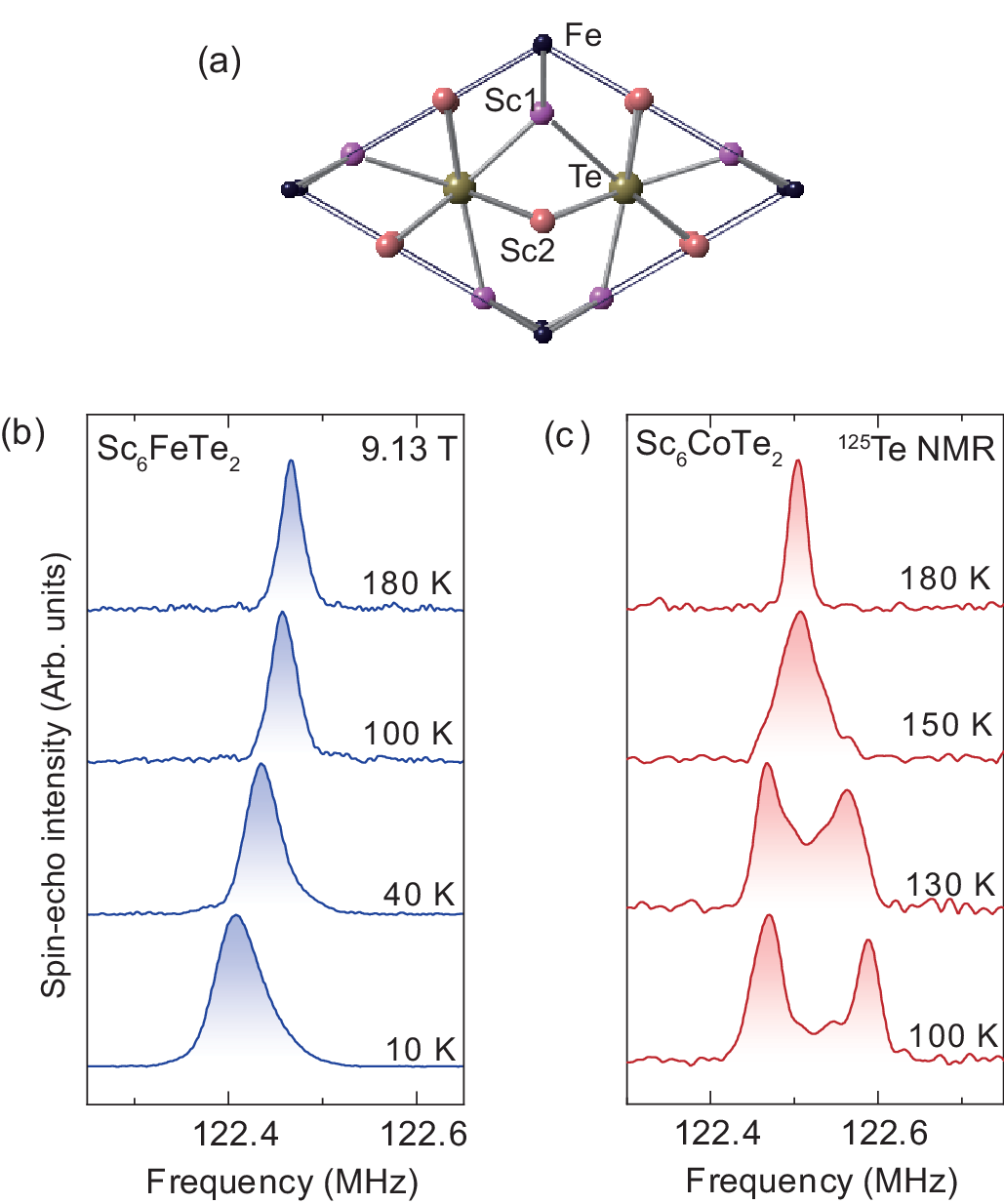}
\caption{\label{spc9T} (a) Crystal structure of Sc$_6$FeTe$_2$. Temperature dependence of $^{125}$Te NMR spectrum for (b) Sc$_6$FeTe$_2$ and (c) Sc$_6$CoTe$_2$ under magnetic field $H_0$ = 9.13 T. }
\end{figure}

A single $^{125}$Te NMR spectrum was observed at high temperatures, as shown in Fig. \ref{spc9T}, consistent with the single Te site of the $P\bar{6}2m$ lattice. The nearly symmetric spectrum reflects the isotropic hyperfine coupling tensor and spin susceptibility. The spectrum of Sc$_6$FeTe$_2$ shifts to a lower frequency and broadens upon cooling. The line shape becomes asymmetric at low temperatures, which implies a distribution of the local spin susceptibility or an inhomogeneous dipole field from magnetic impurities. 

\begin{figure}
\includegraphics[width=7.5cm,keepaspectratio]{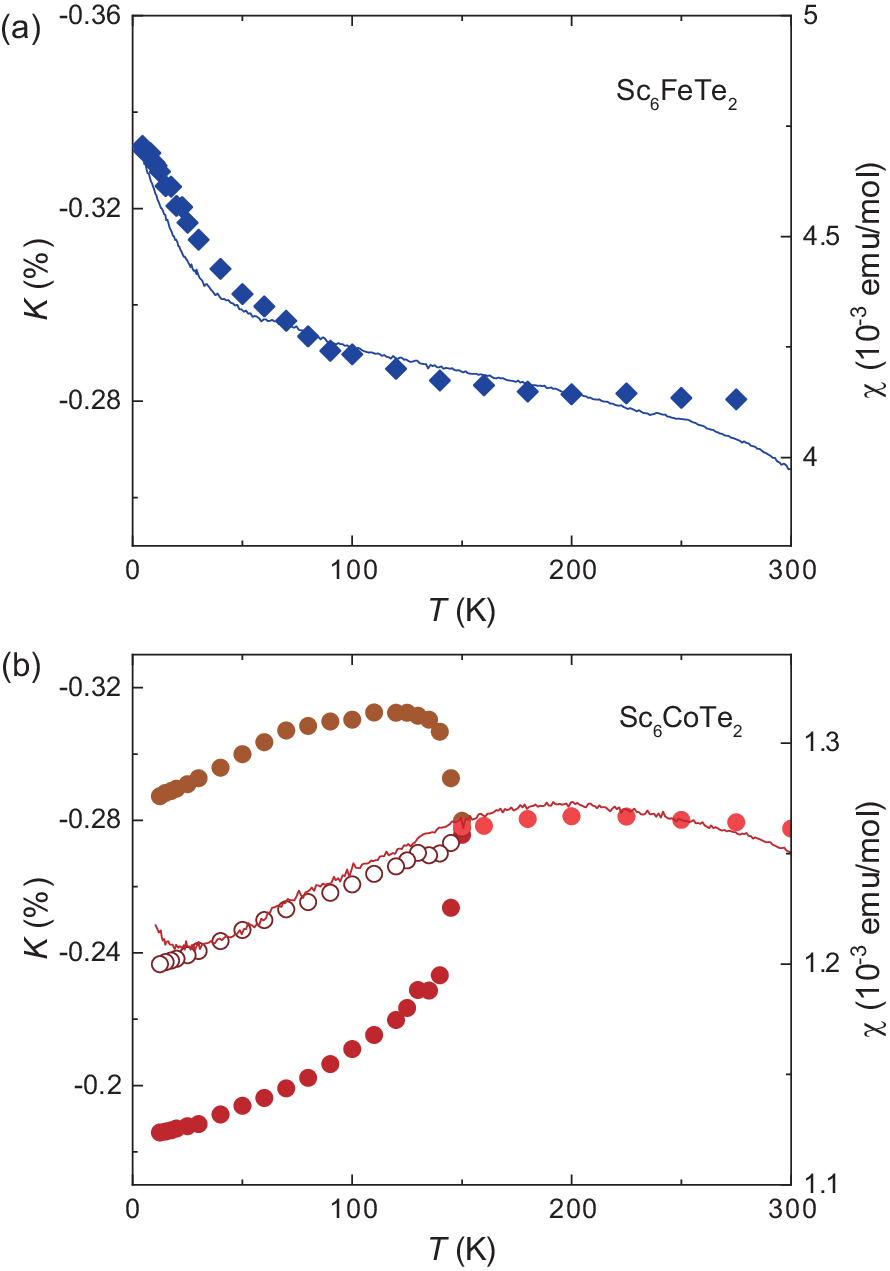}
\caption{\label{K} $^{125}$Te Knight shift $K$ (the left-hand axis) and the bulk magnetic susceptibility $\chi$ (solid curves, the right-hand axis) against temperature $T$ for (a) Sc$_6$FeTe$_2$ and (b) Sc$_6$CoTe$_2$. Open symbols in (b) are the averaged $K$ for two Te sites below $T^*$. Note that both vertical axes are expanded for clarity. }
\end{figure}

The Knight shift $K$ determined from the peak position scales to $\chi$ below 200 K, as shown in Fig. \ref{K}(a). As the bulk spin susceptibility $\chi$ increases upon cooling (a solid curve on the right-hand axis), the negative Knight shift represents the negative hyperfine coupling constant due to the predominant core polarization of $s$-electron spins. The linearity of the $K -\chi$ plot yields the hyperfine coupling constant $H_{\rm hf} = -0.8(2)$ T/$\mu_{\rm B}$ for Sc$_6$FeTe$_2$. The detail of the $K$-$\chi$ analysis is described in the Appendix \ref{appendix}. 

For Sc$_6$CoTe$_2$, the $^{125}$Te NMR spectrum is also narrow at high temperatures, as shown in Fig. \ref{spc9T}. In contrast to Sc$_6$FeTe$_2$, the spectrum starts to broaden around $T^{*} \sim$ 150 K and splits into two at lower temperatures. The splitting indicates a doubling of the Te site by breaking the mirror symmetry in the unit cell. The transition occurs continuously, as shown in Fig. \ref{spc9T}(b), where $K$ defined by each spectral peak shows the contrasting temperature dependence below $T^*$: one Te spectrum shifts to a higher frequency just below $T^*$, while another shifts to a lower side. 

For $^{125}I$ = 1/2 without an electric quadrupole interaction, a spectral splitting comes purely from a magnetic origin, e.g. the difference in the local density of states between two Te sites. An increase in the averaged shift of two spectra below $T^*$ represents a suppression of $\chi$ for the negative hyperfine coupling constant. The scaling between the averaged $K$ yields $H_{\rm hf} = -3.4(3)$ T/$\mu_{\rm B}$. Since the density of states at the Fermi level, $D(E_{\rm F})$, is governed by the transition metal $M$ and the scandium sites in the band structure calculation \cite{Shinoda2023}, the hyperfine coupling at the Te sites would come from the transferred interaction with surrounding ions. The observed structural transition in Sc$_6$CoTe$_2$ implies a structural instability for a symmetry lowering possibly due to the orbital-lattice coupling of transition metals. 

A weak thermal variation of $K$ and spin susceptibility is attributed to the dependence of $D(E_{\rm F})$ averaged over the energy range of $k_{\rm B}T$. The spin susceptibility proportional to $D(E_{\rm F})$ is enhanced (or suppressed) when $E_{\rm F}$ is located close to the ridge (or valley) of the density of states, consistent with the band calculation in Sc$_6$FeTe$_2$ (Sc$_6$CoTe$_2$), respectively \cite{Shinoda2023}. 

\begin{figure}
\includegraphics[width=8.6cm,keepaspectratio]{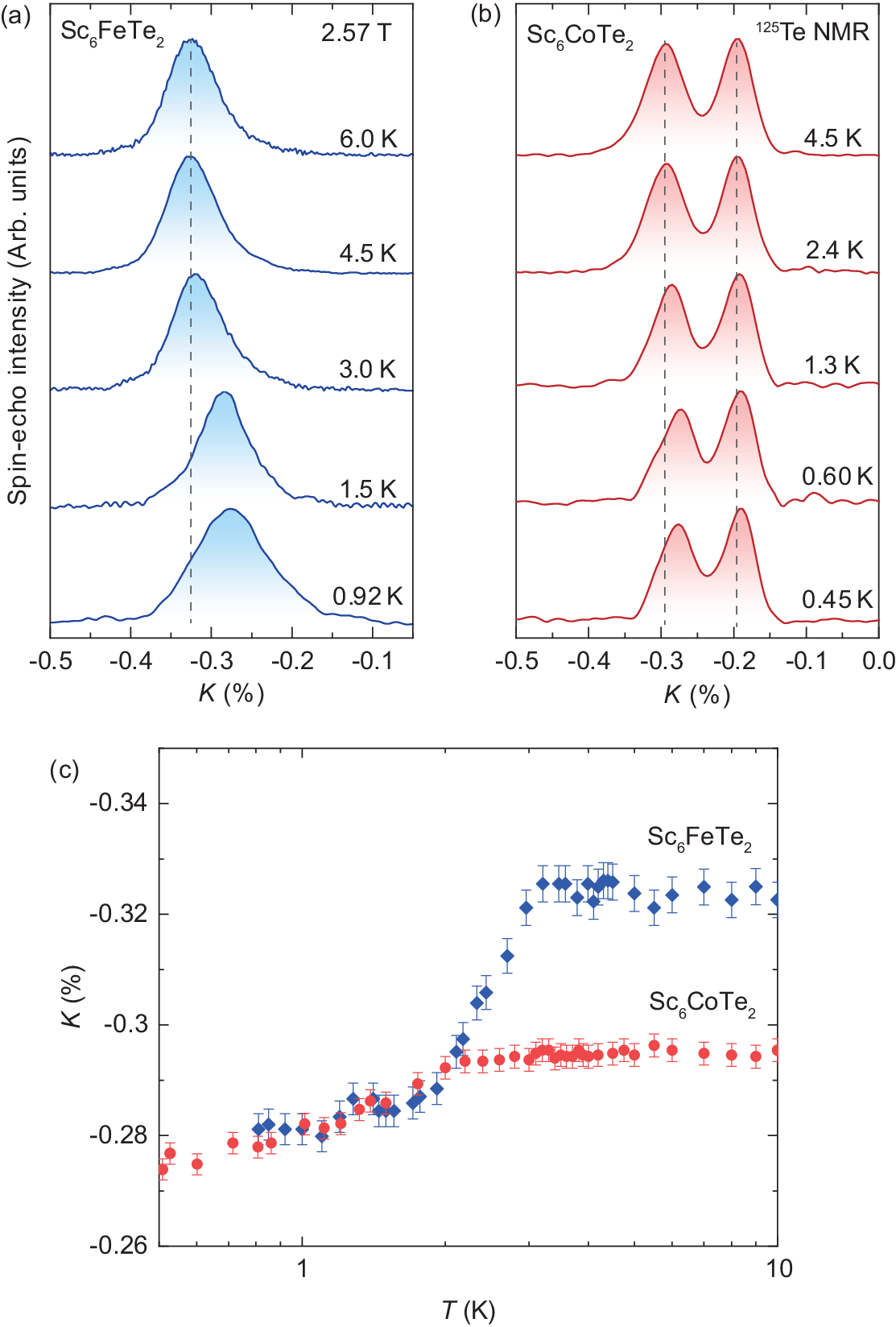}
\caption{\label{spc2T} Temperature dependence of $^{125}$Te NMR spectra for (a) Sc$_6$FeTe$_2$ and (b) Sc$_6$CoTe$_2$ across $T_c$ at $H_0$ = 2.57 T. Dashed lines indicate the peak positions above $T_c$. (c) $K$ obtained from the peak position of the NMR spectrum. Only the lower frequency shift is plotted in Sc$_6$CoTe$_2$.}
\end{figure}

At low fields below $H_{c2}$, the Knight shift gives the spin susceptibility in the superconducting phase. We measured $^{125}$Te NMR spectra across $T_{\rm c}$ at \textit{H} = 2.57 T well below $H_{c2}$. $T_c$ is obtained as 3.4 and 2.5 K for Sc$_6$FeTe$_2$ and Sc$_6$CoTe$_2$, respectively at $H = 2.6$ T in the resistivity measurement \cite{Shinoda2023}. As shown in Fig. \ref{spc2T}, the peak frequency of the spectrum is almost independent of the temperature in the normal state below 10 K and starts to shift in a positive direction below $T_c$ for Sc$_6$FeTe$_2$. For a negative hyperfine coupling constant, the spin susceptibility is suppressed below $T_c$, as expected in spin-singlet superconductors \cite{Yosida1958}.

For Sc$_6$CoTe$_2$, one of the spectra at the lower frequency exhibits a positive shift below $T_{\rm c}$, as shown in Fig. \ref{spc2T}(b). In contrast, the spectrum for the higher frequency exhibits a tiny shift. The site-dependent shift rules out the extrinsic origin due to the Meissner effect and reflects the difference in the spin density between two Te sites, which participates in the superconducting pairing. For the negative hyperfine coupling, the positive Knight shift also indicates a decrease in spin susceptibility below $T_{\rm c}$. The residual value at the lowest temperature can be attributed to the paramagnetic component arising from spin-triplet pairing or a residual density of states. 

The Knight shift generally consists of the spin part $K_{\rm s}$ and the orbital part $K_{\rm orb}$. Assuming Korringa's relation described in the following, the orbital part $K_{\rm orb}$ is evaluated from the $K$-$(T_1T)^{-0.5}$ plot as $K_{\rm orb}$ = 0.02\% for Sc$_6$FeTe$_2$ and $K_{\rm orb}$ = 0.05\% for Sc$_6$CoTe$_2$ (See Appendix \ref{appendix}). The difference of $K$ between the normal and superconducting states is tiny: $\Delta K = 0.04$\% for Sc$_6$FeTe$_2$ and $\Delta K = 0.02$\% for Sc$_6$CoTe$_2$ [Fig. \ref{spc2T}(c)], which are one order smaller than those expected in the spin-singlet state. The spectral broadening below $T_{\rm c}$ can be attributed to a magnetic impurity and the penetration of vortex cores. 

\begin{figure}		
\includegraphics[width=6.5cm,keepaspectratio]{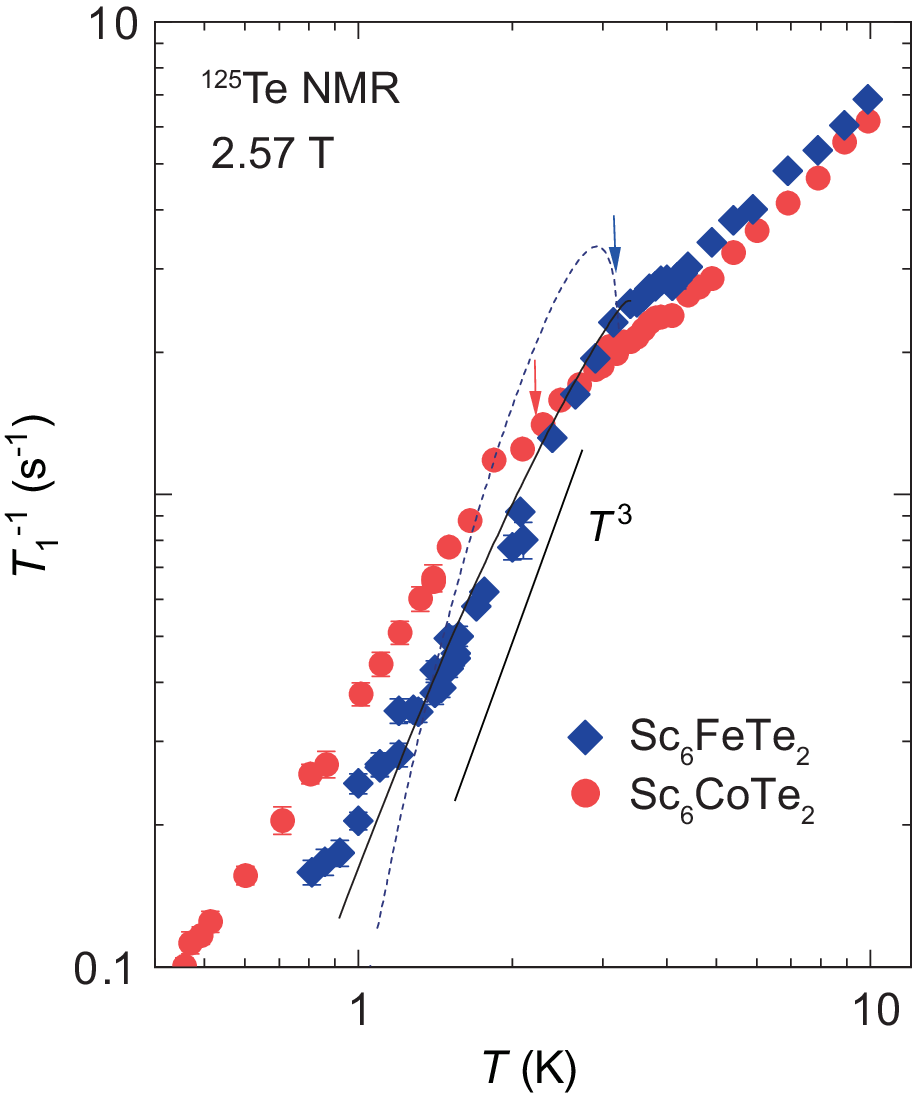}
\caption{\label{T1}Temperature dependence of $1/T_1$ measured at 2.57 T. Arrows indicate the superconducting transition temperature $T_c$ determined by the zero resistance \cite{Shinoda2023} and the {\it in-situ} ac susceptibility measurement. The dotted and solid curves are numerical calculations fro conventional full gap \cite{Hebel1957} and line-node gap superconductor. The solid line is a guide of the power law ($\sim T^3$).}
\end{figure}

The superconducting gap structure was investigated by $1/T_1$ in a magnetic field of 2.57 T, as shown in Fig. \ref{T1}. $1/T_1$ linearly decreases with temperature in the normal state and shows a steep drop below $T_c$ = 3.3 and 2.2 K for Sc$_6$FeTe$_2$ and Sc$_6$CoTe$_2$, respectively. The coherence peaks expected in $s$-wave superconductors were not observed just below $T_c$. Instead, $1/T_1$ follows the power law $\propto T^n$ ($n \simeq 3$) rather than the exponential function expected in a full-gap superconductor. The result is consistent with the presence of line nodes in the quasiparticle excitation gap. 

\begin{figure}
\includegraphics[width=8cm,keepaspectratio]{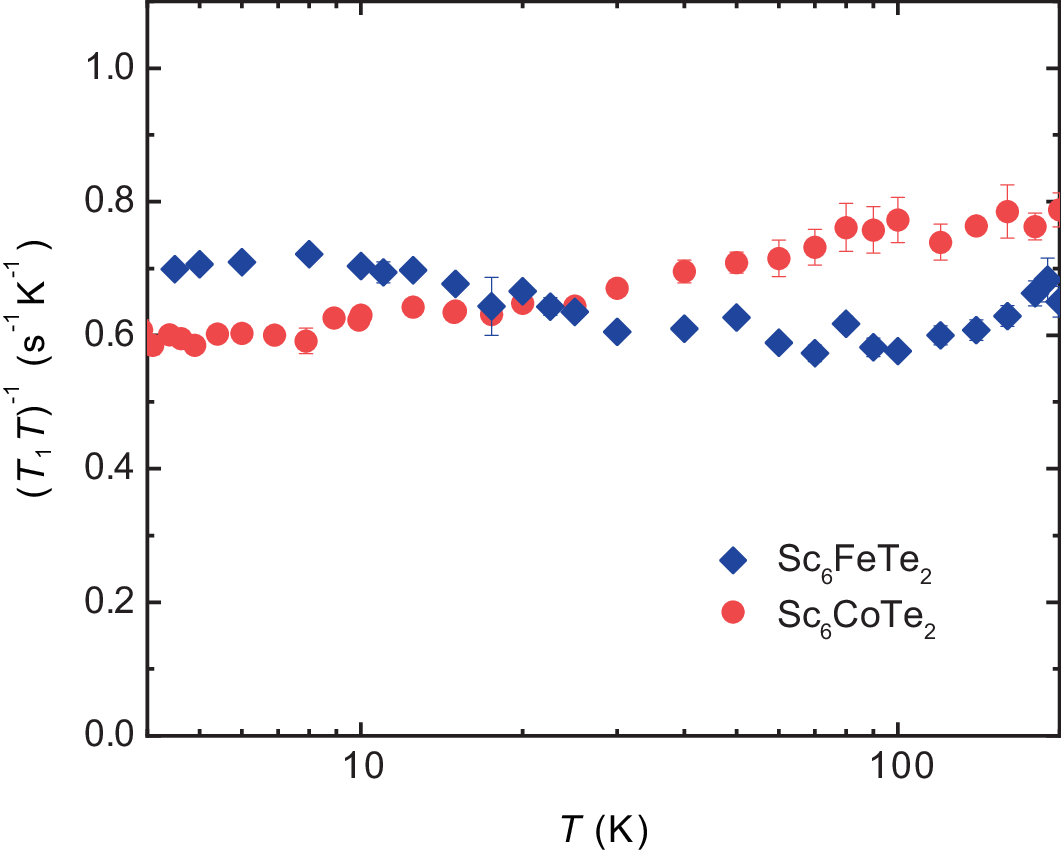}
\caption{\label{T1T} Nuclear spin-lattice relaxation rate divided by temperature, $(T_1T)^{-1}$, on Sc$_6$FeTe$_2$ and Sc$_6$CoTe$_2$ at $H$ = 9.13 T. We obtained $T_1$ for the integrated spin-echo intensity of the two Te sites in Sc$_6$CoTe$_2$ below $T^*$. No indication of a magnetic transition is observed at 150 K, where the NMR spectrum starts to split. }
\end{figure}

In standard metals, $1/T_1T$ is proportional to the square of $D(E_{\rm F})$. As shown in Fig. \ref{T1T}, $1/T_1T$ increases slightly or decreases upon cooling for Sc$_6$FeTe$_2$ and Sc$_6$CoTe$_2$, respectively. The behavior is similar to those of $\chi$ and $K$ proportional to $D(E_{\rm F})$ in Fig. \ref{K}. Therefore, Korriga's relation, $1/T_1T \propto K_{\rm s}^2$, holds in both compounds (See also the Appendix \ref{appendix}). That is, the ratio $1/T_1TK_s^2$ takes the universal constant called the Korringa constant $\mathcal{K}_\alpha = \frac{\hbar}{4\pi k_{\mathrm{B}}} \left( \frac{\gamma_\mathrm{e}}{\gamma_\mathrm{n}} \right)^2 \frac{1}{T_1TK_{\mathrm{s}}^2}$, where $\gamma_e$ is the electron gyromagnetic ratio. $\mathcal{K}_\alpha$ would be progressively enhanced or suppressed against temperature in the presence of antiferromagnetic or ferromagnetic fluctuations, respectively. Here we obtained temperature-independent $\mathcal{K}_\alpha$ = 1.0 for Sc$_6$FeTe$_2$ and 1.3 for Sc$_6$CoTe$_2$, suggesting a negligible effect of spin fluctuations in the normal state. 

In a weakly correlated metal, the superconducting order parameter would have a $s$-wave nature, as anticipated from the BCS theory mediated by phonons, whereas it can be anisotropic in the presence of strong electron-electron and electron-phonon coupling. In Sc$_6M$Te$_2$, superconductivity induced by strong electron-phonon coupling has been recently proposed on the basis of the {\it ab initio} calculation \cite{Jiang2024}, where the anharmonic low-frequency phonon (rattling) mode plays a crucial role in superconductivity. Strong coupling superconductivity was also expected from the specific heat behavior across $T_c$ in Sc$_6$FeTe$_2$ \cite{Shinoda2023}, which is compatible with the present NMR results showing the anisotropic order parameter. The strong electron-phonon coupling, if any, can be observed as an enhancement of $1/T_1T$, as is known in KOs$_2$O$_6$ having a rattling phonon mode \cite{Yoshida2007}. To uncover the phonon dynamics, we have measured $1/T_1$ at the $^{45}$Sc sites with electric quadrupole interaction for Sc$_6$FeTe$_2$. It also follows Korringa's law without the enhancement of $1/T_1T$ in the normal metallic state. 

Note that the coherence peak is sensitively suppressed in the presence of magnetic field and impurities \cite{Masuda1971, Choi1995, Mitrovic2006, Yang2015}. The gap structure and topology can change by impurities, as discussed in Fe pnictide superconductors \cite{Mizukami2014}. For Sc$_6$FeTe$_2$, the Sommerfeld coefficient $\gamma_0$ remains finite even under zero magnetic field and increases with magnetic field, implying a small amount of extrinsic normal-state domain around defects \cite{Shinoda2023}. We observed a slight suppression of the stretched exponent well below $T_c$, indicating the residual normal state under the magnetic field. Although further low-field experiments are desired, we have not succeeded in detecting the NMR signal of the superconducting phase at lower fields due to the Meissner shielding against the rf pulse. 

In Sc$_6M$Te$_2$, the band splitting due to the absence of the inversion symmetry would be too small to admix the spin-singlet and spin-triplet pairing \cite{Shinoda2023}. If the spin-triplet component exists, the spin susceptibility depends on the magnetic field direction: it is independent of temperature in a magnetic field parallel to the $d$-vector of spin-orbit coupling, while it decreases below $T_{\rm c}$ for a field normal to the $d$ vector. Taking into account the $d$ vector parallel to the $c$ axis, spin susceptibility would decrease along the in-plane direction, leading to a positive change below $T_{\rm c}$. Then, the NMR spectrum for a powder sample is distorted into the axially anisotropic powder pattern in the superconducting phase, which does not match our experimental result. For confirmation of parity mixing, it would be necessary to measure the anisotropy of the spin susceptibility on a single crystal. 

\section{Conclusion}
In conclusion, the symmetry of the superconducting order parameter was investigated on the noncentrosymmetric Sc$_6M$Te$_2$ without inversion symmetry through $^{125}$Te NMR measurements. The spin susceptibility obtained from the Knight shift is suppressed below $T_c$, consistent with the spin-singlet pairing, while the large residual value implies parity mixing of the spin-triplet and singlet components due to spin-orbit coupling. The temperature dependence of $1/T_1$ provides evidence for the nodal gap in the quasiparticle excitation. The possibility of spin-singlet and triplet superconductivity should be carefully considered through lower-field experiments with a high-quality single crystal. 

\section*{Acknowledgements}
We are thankful for useful discussion with Y. Yamakawa and H. Kontani. We acknowledge support from grants-in-aid in scientific research by JSPS (No.JP19H05824, No. 22H05256, No. 22K03510, No. 23H04025, and No. 24H00954).

\appendix

\section{Evaluation of orbital Knight shift}
\label{appendix}

\begin{figure}
\includegraphics[width=8.8cm,keepaspectratio]{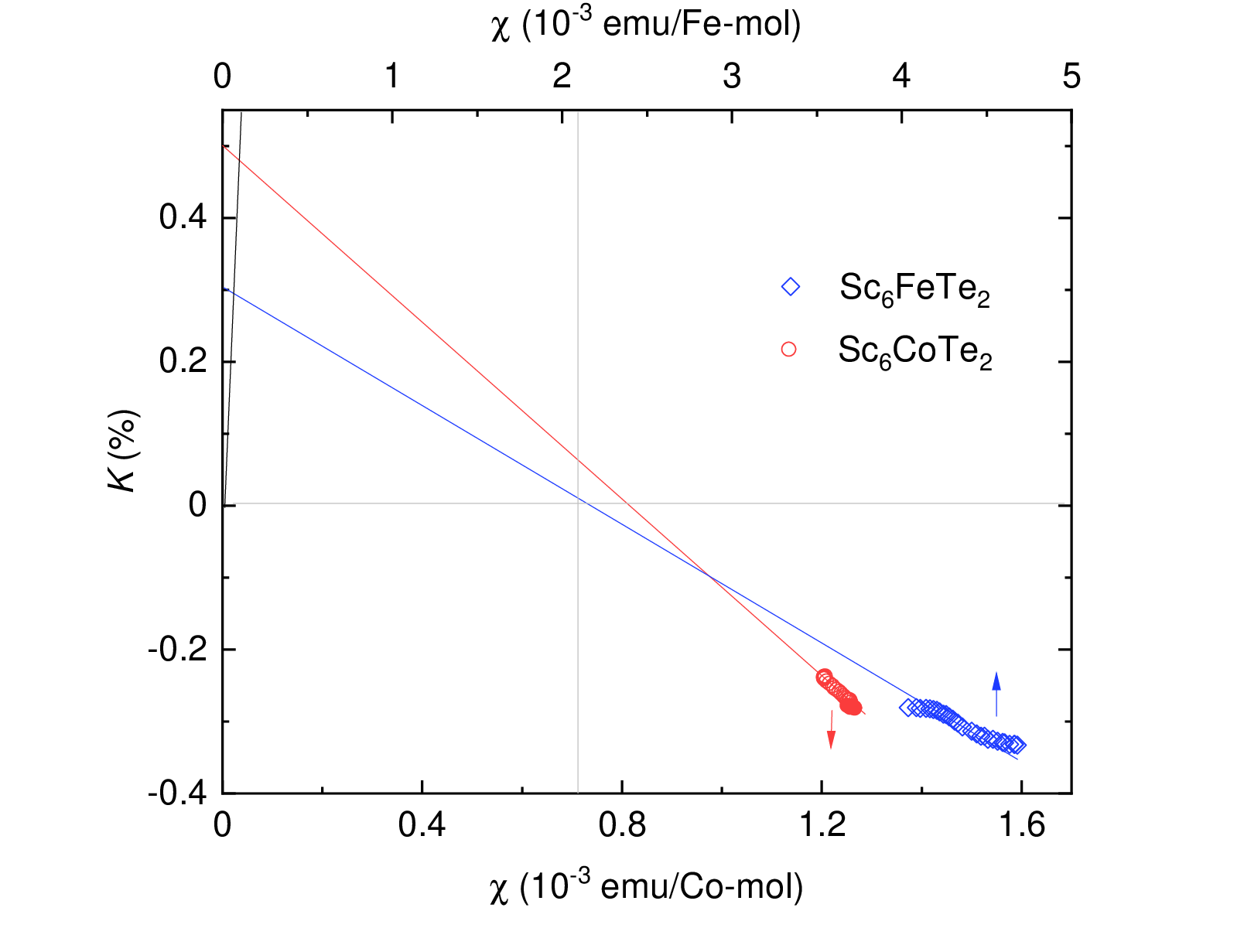}
\caption{\label{KChi} Knight shift vs bulk magnetic susceptibility as an implicit function of temperature for Sc$_6$FeTe$_2$ and Sc$_6$CoTe$_2$ at $H$ = 9.13 T. $K$ was defined as an average shift for two Te sites in Sc$_6$CoTe$_2$ below $T^*$. The black line is the calculated orbital shift. }
\end{figure}

As described above, the NMR Knight shift consists of spin and orbital parts. The spin part $K_{\rm s}$ can depend on temperature and be proportional to spin susceptibility. The orbital part $K_{\rm orb}$ governed by the Van-Vleck orbital susceptibility is independent of temperature and subtracted from the $K$ versus $\chi$ plot in Fig. \ref{KChi}. Good linearity is observed in the temperature range of 100--180 K and 20--200 K for Sc$_6$FeTe$_2$ and Sc$_6$CoTe$_2$, respectively. The linearity yields the hyperfine coupling constant, as described in the main text. The crossing points between spin and orbital linearity give the constant offset $K_{\rm orb}$ = 0.29\% (Fe) and 0.48\% (Co). 

\begin{figure}
\includegraphics[width=8.8cm,keepaspectratio]{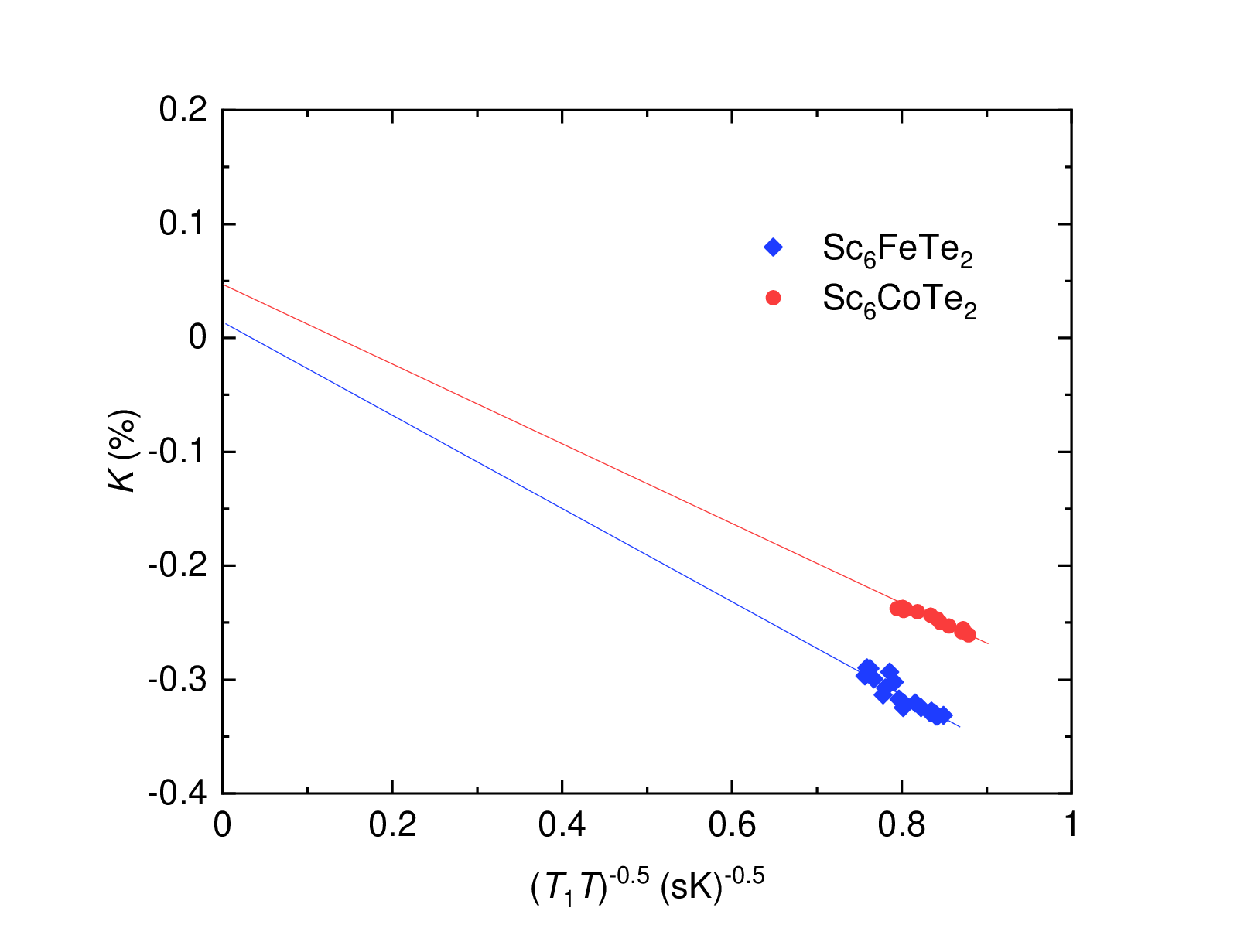}
\caption{\label{T1TK} Knight shift $K$ plotted against the square of the nuclear spin-lattice relaxation rate divided by temperature, $(T_1T)^{-0.5}$, on Sc$_6$FeTe$_2$ and Sc$_6$CoTe$_2$ at $H$ = 9.13 T. }
\end{figure}

In Fermi liquid metal without spin fluctuations, $K_{\rm s}$ and $(T_1T)^{-1}$ are proportional to $D(E_{\rm F})$ and $D(E_{\rm F})^2$, respectively. Then $K_{\rm s}$ is linear to $(T_1T)^{-0.5}$ as an implicit function of temperature. Here the orbital contribution in $(T_1T)^{-1}$ would be negligible for Te sites without orbital degeneracy. The extrapolation in the $K$-$(T_1T)^{-0.5}$ linearity can be another way to evaluate $K_{\rm orb}$ \cite{Harada2012}, which should be consistent with the $K$-$\chi$ plot. 

We find a good linear relation between $K$ and $(T_1T)^{-0.5}$, as shown in Fig. \ref{T1TK}. We obtained $K_{\rm orb} = 0.02$\% and 0.05\%, which is largely different from the $K$-$\chi$ plot. The discrepancy may come from the significant constant offset of the bulk magnetization, as observed in Fe and Co pnictide superconductors \cite{Ahilan2014, Cui2015}. In the present case, it can be attributed to the small amount of magnetic impurities.

\bibliography{bib}

\end{document}